# On Periodic Node Deployment in Wireless Sensor Networks: A Statistical Analysis

Abhishek Sinha[1], Swagatam Das[1], and Athanasios V. Vasilakos[2]

[1]Dept. of Electronics and Telecommunication Eng, Jadavpur University, Kolkata, India.
[2]Dept. of Computer and Tele Eng, University of Western Macedonia, Greece

abhishek_sinha_ju@yahoo.com, swagatamdas19@yahoo.co.in, vasilako@ath.forthnet.gr

**Abstract:** Rapid progress made in the field of sensor technology, wireless communication, and computer networks in recent past, led to the development of wireless Ad-hoc sensor networks, consisting of small, low-cost sensors, which can monitor wide and remote areas with precision and liveliness unseen to the date without the intervention of a human operator. This work comes up with a stochastic model for periodic sensor-deployment (in face of their limited amount of battery-life) to maintain a minimal node-connectivity in wireless sensor networks. The node deployment cannot be modeled by using results from conventional continuous birth-death process, since new nodes are added to the network in bursts, i.e. the birth process is not continuous in practical situations. We analyze the periodic node deployment process using discrete birth-continuous death process and obtain two important statistical measures of the existing number of nodes in the network, namely the mean and variance. We show that the above mentioned sequences of mean and variances always converge to finite steady state values, thus ensuring the stability of the system. We also develop a cost function for the process of periodic deployment of sensor nodes and minimize it to find the optimal time ($\tau$) and optimum number of re-deployment ($q$) for maintaining minimum connectivity in the network.

# 1. Introduction

An ad-hoc wireless sensor network (WSN) consists of a number of sensors spread across a geographical area. Each sensor has wireless communication capability and some level of intelligence for signal processing and networking of the data. The development of WSNs was originally motivated by military applications such as battlefield surveillance. However, they are now used in many industrial and civilian application areas, including industrial process monitoring and control, machine health monitoring, environment and habitat monitoring, healthcare applications, home automation, and traffic control [1 – 4]. Because of its high practical applications, Massachusetts Institute of Technology (MIT) ranked this network as one of the ten burgeoning technologies that has great influence on people's day-to-day life, in 2003 [5]. Past few years have witnessed a great advance in the research dedicated to sensor networks, including design issues related to the physical and media access layers [6 - 8] and routing and transport protocols [9 - 13]. Localization and positioning applications of wireless sensor networks are discussed in works like [14 - 18]. A few excellent surveys on the present state-of-the-art research on sensor networks can be traced in [19 - 23].

Since sensors in an ad-hoc network may spread in an arbitrary manner, one of the fundamental issues in a wireless sensor network is the *coverage problem*. In general, this reflects how well an area is monitored or tracked by sensors. One of the main objectives of the coverage problem is to prolong network life-time. Besides this objective, the coverage problem has many variants based on the other constraints of networks and objects to be covered. Many algorithms, including centralized, distributed and localized coverage, have been proposed in [24 –32]. Most of these algorithms focus on a mechanism to geometrically organize the sensors into a number of clusters such that each cluster can completely cover a portion of the region of interest maintaining connectivity.

The actual coverage problem can be thought of as a combination of two hierarchical sub-problems: at higher level the problem of Geometrical distribution of nodes for optimum coverage, network lifetime and connectivity and at a lower level, the problem of determining the rate at which nodes should be re-deployed in the network in order to maintain an optimal connectivity, in face of the continuous death of the sensor nodes (may be due to damned environment, limited battery lifetime etc.). This work presents a humble attempt to contribute in the latter context.

Energy efficiency is one of the most critical issues in WSNs. With the current available technology, sensors are battery powered and have a limited weight. These characteristics globally affect the application lifetime. As the WSN remains operational, due to energy consumption for sending and relaying the environmental data, sensor nodes gradually run short of battery and ultimately they become un-operational. So to maintain connectivity among the nodes, it is instructive to periodically deploy the sensor nodes on the operational area. Now if we deploy any number of nodes at any time we may run into problems either of:

1) Overflowing the network with redundant amount of nodes, thus making the system potentially unstable and unnecessarily increasing the network maintenance cost; or,
2) Sacrificing the connectivity of the network for some time interval for lack of adequate number of sensor nodes.

To address this problem this work first aims to build a mathematical model of the periodic node deployment phenomena from scratch, and then subjecting then we attempt to minimize the average network maintenance cost. Note that in [31], the authors attempted to describe the change of sensor nodes using the birth-death model [33, 34] and Markov chain [35]. However, they assumed continuous birth (i.e. addition of new nodes to the network); where in reality the birth or node addition in the network is in bursts i.e. discrete. We take this into account and build an independent statistical model of node deployment in

WSNs. Based on our discrete birth-continuous death model for node deployment we derive two important statistical measures of the existing number of sensor nodes in the steady state, namely the mean ($\mu$) and variance ($\sigma^2$) We analytically prove that the above mentioned sequences of mean and variances converge, thus ensuring the stability of the system. We also develop a cost function for the process of periodic deployment of nodes and minimize it to find the optimum period τ for the deployment to maintain minimum connectivity. We assume that the sensor nodes are distributed randomly with a uniform density, and all sensors have the same range radius $R$. Note that here we study the coverage of the sensor nodes on a statistical basis, and not from a geometrical perspective. We deploy the sensor nodes uniformly thus eliminating any geometrical contribution of individual sensor-positions whatsoever. We define the coverage performance of the network as:

$$C_{coverage} = K \cdot \frac{\pi R^2}{D} \propto \frac{K}{D}, \tag{1}$$

where

$K$ = number of existing sensor nodes,

$D$ = area of operation,

$R$ = coverage radius of individual sensor nodes

Thus $K/D$ equals the number of existing sensor nodes per unit square area, and it is a direct measure of coverage for uniform sensor distribution. In the present work, this is precisely the statistical measure of this quantity which we have calculated theoretically, ensuring the coverage of the sensors. Unless otherwise stated, in the subsequent discussion, all the numbers referring to the number of sensor nodes denotes numbers per unit square area.

## 2. A Mathematical Model for Node Deployment

First we consider the case in which $t \in (0, \tau)$. In this open interval no node is added to the sensor network. So we have $\lambda_n = 0$ for all $n = 0, 1, 2, \cdots$. The nodes are added to the network only at the time instants $t = 0$ Suppose, it is known deterministically that there were $j$ number of nodes at the beginning of the interval.

Let us denote the probability that at time instant $t \in (0, \tau).$, the network consists of $n$ number of sensor nodes by $p_n(t)$. Also, let the number of deaths (due to energy exhaustion, destruction caused by damned environment etc.) per unit time, when the existing number of sensor nodes is $n$, be given by $\mu_n$. Then the probability that at time $t + \Delta t$, the number of existing sensor nodes remains $n$ is given by $p_n(t + \Delta t)$. Then assuming continuous time Markov model for the process, we have,

$p_n(t + \Delta t) =$ (Probability that there were $(n+1)$ nodes at time $t$) * (Probability of one death in time interval $(t, t + \Delta t)$ + Probability that there were $n$ nodes at time $t$) * (Probability of no death in time interval $(t, t + \Delta t)$ )) + $O(\Delta t^2)$

(Here the term $O(\Delta t^2)$ represents the probability of more than one death in the small time interval $\Delta t$ )

or, $p_n(t + \Delta t) = p_{n+1}(t).\mu_{n+1}.\Delta t + p_n(t).(1 - \mu_n.\Delta t) + O(\Delta t^2)$

or, $\dfrac{p_n(t + \Delta t) - p_n(t)}{\Delta t} = p_{n+1}(t).\mu_{n+1} - p_n(t).\mu_n + O(\Delta t)$

Taking limit as $\Delta t \to 0$ on both sides, we get the desired result as:

$$\dfrac{dp_n}{dt} = p_{n+1}(t).\mu_{n+1} - p_n(t).\mu_n \qquad (2)$$

Taking the number of existing sensor nodes as the current state of the system, we assume that the death rate at state $n$ is linearly proportional to the state $n$, i.e. $\mu_n = n.\mu$ (i.e. the death rate is directly proportional to the existing number of nodes), the rate of death in unit interval being $\mu$ per individual. Taking this into account, we have the following differential-difference equations for $p_n(t) = \Pr\{X(t) = n\}$

$$\dfrac{dp_0}{dt} = \mu.p_1(t) \qquad (3a)$$

$$\dfrac{dp_n}{dt} = (n+1).\mu.p_{n+1}(t) - n.\mu.p_n(t) \ ; 0 < n < j \qquad (3b)$$

$$\dfrac{dp_j}{dt} = -(j+1).\mu.p_j, \qquad (3c)$$

and $\dfrac{dp_n}{dt} = 0$ otherwise.

Let us define the probability generating function $P(s,t) = \sum\limits_{n=1}^{\infty} p_n(t).s^n$ from standard probability theory [36, 37]. Taking partial derivatives w. r. t. $s$ and $t$,

$$\dfrac{\partial P}{\partial s} = \sum\limits_{n=1}^{\infty} n p_n(t) s^{n-1}$$

and $\dfrac{\partial P}{\partial t} = \sum\limits_{n=0}^{\infty} p'_n(t) s^n$

Multiplying equation (3b) by $s^n$, summing over $n = 1, 2 \ldots \infty$ and finally adding equation (3a) and (3c) with it, we get:

$$\frac{\partial P}{\partial t} = -\mu s \frac{\partial P}{\partial s} + \mu \frac{\partial P}{\partial s}$$

$$= \mu.(1-s).\frac{\partial P}{\partial s} \qquad (4)$$

This is the partial differential equation satisfied by $P(s,t)$, subject to the boundary condition:

$$P(s,0) = \sum_{n=0}^{\infty} p_n(0).s^n = s^j, \qquad (5)$$

since initially at time $t = 0$, the network consists of $j$ nodes i.e. only $p_j(0)$ is 1 and all other $p_n(0) = 0$ for $n \neq j$. Now our task will be to solve equation (4) subject to the boundary condition given in (5).

It is not difficult to guess the solution of the equation which has a simple form as follows

$$P(s,t) = [1-(1-s).e^{-\mu.t}]^j \qquad (6)$$

It readily follows that (6) indeed satisfies both the PDE and its associated boundary condition given in (4) and (5) respectively. Based on this solution, we enunciate the following theorems:

**Theorem 1**: Expected number of sensor nodes at time $t = t_m$ is given by:

$$\frac{\partial P}{\partial s}\bigg|(s=1, t=t_m)$$

**Proof**: Expected number of sensor nodes at time $t = t_m$ (with initial $j$ number of nodes) is given by

$$<n>_j = \sum_{n=0}^{\infty} n.p_n(t_m).$$

Now as we have seen, $\frac{\partial P(s,t)}{\partial s} = \sum_{n=0}^{\infty} n.p_n(t).s^{n-1}$

So $\frac{\partial P}{\partial s}\bigg|(s=1, t=t_m) = \sum_{n=0}^{\infty} n.p_n(t_m) = <n>_j$ (Proved)

**Theorem 2**: The expected number of sensor nodes at time $t = t_m$ (with $j$ initial number of nodes) given by

$$<n>_j = je^{-\mu.t_m}$$

**Proof**: As we have seen earlier from (6),

$$P(s,t) = [1-(1-s)e^{-\mu.t}]^j$$

From theorem (1), $<n>_i = \frac{\partial P}{\partial s}\bigg|(s=1, t=t_m)$

$$= \frac{\partial}{\partial s}[1-(1-s).e^{-\mu.t}]^j\bigg|_{s=1, t=t_m}$$

$$= j.e^{-\mu.t_m} \qquad \text{(Proved)}$$

**Corollary 1**: If the initial number of nodes $j$ be a random variable with expected value $<j>$, then expected number of sensor nodes after time $t = t_m$ is given by:

$$<n> = <j>.e^{-\mu.t_m}$$

**Proof**: by definition, $<n> = \sum_{j=0}^{\infty} <n>_j .p_j(0)$

$$= \sum_{j=0}^{\infty} j.e^{-\mu.t_m}.p_j(0)$$

$$= e^{-\mu.t_m}.\sum_{j=0}^{\infty} j.p_j(0)$$

$$= <j>.e^{-\mu.t_m} \quad \text{(Proved)}$$

Now based on the above results, we address the problem of stability and convergence of the number of nodes by considering periodic node deployment at a time interval of $\tau$.

## 3. Stability and Convergence of Expected Number of Nodes and their Variance

In this section we show that, taking into account the continuous death process as described in the previous section, if new sensor nodes are periodically deployed in the WSN (i.e. say $q$ nodes added in bursts after every $\tau$ seconds), then the expected number of nodes in the network converges to a steady state value as time $t \to \infty$. Below we first outline the steps of a computational procedure that will be used to investigate the convergence of expected number of nodes $<n>$ in the network at any future instant. All the superscripts in the material denote suffixes and not exponents.

*Step 1*: Solve the difference-differential equation given in (3a), (3b), and (3c) subject to the initial condition $n = j$ at $t = 0$. Let the solution be

$$p_n^j(t).$$

*Step 2*: Now calculate the probability of being in state $n$ after time $\tau$ (just before the next deployment), which is equal to: $P^1(n) = p_n^j(\tau)$

*Step 3*: Having calculated $P^1(n)$, recursively compute $P^2(n), P^3(n)$, (i.e. the probability of being in state '$n$' just before 2nd, 3rd, …. deployments )… using the recursive formula for $P^{(k)}(n)$ derived later (in Theorem 3). Remember that all the values of $P^2(n)$, $P^3(n)$ etc. denotes the probabilities just before the next deployment.

*Step 4*: Having calculated $P^k(n)$ s, it is easy to calculate the expected value of the surviving nodes ($m^{(k)}$) and their variance ($Var^{(k)}$)) just before the next node deployment at $k\tau - 0$ by the simple formula:

$$m^{(k)} = \sum_n n.P^k(n)$$

$$Var^{(k)} = \sum_n (n - m^{(k)})^2 P^{(k)}(n)$$

Let us denote the number of existing sensor nodes just before the $k^{th}$ deployment by the discrete time stochastic process $X^{(k)}$. Then it can be said that the probability distribution of the stochastic process $X^{(k)}(n)$ is given by $P^{(k)}(n)$.

Recall that our objective is to investigate the steady state behavior of $m^{(k)}$ and $Var^{(k)}$. Now from step 4, we show that $m^{(k)}$ and $Var^{(k)}$ indeed converge irrespective of the value of $q$, $\tau$ and initial number of nodes (*j*). In what follows, we adopt the conventions that:

1. $$^pC_m = 0 \quad ; \text{ if } m < 0 \text{ or } p < m$$
   $$= \frac{p!}{m!(p-m)!} \quad ; \text{ otherwise}$$

2. A running index (such as '*n*' or '*i*') of a summation takes all possible integral values starting from zero.

3. A notation like $x - 0$ denotes an infinitesimally close point to $x$ but less than $x$.

From equation (6) we have, $P(s,t) = [1 - (1-s)e^{-\mu.t}]^j$

$$= [(1 - e^{-\mu.t}) + se^{-\mu.t}]^j \tag{7}$$

From the definition of $P(s,t)$, $p_n(t)$ equals the coefficient of $s^n$ in the expansion of $P(s,t)$ in powers of *s*. So expanding (7) binomially in powers of *s*, we get,

$$p_n^j(t) = {}^jC_n(1 - e^{-\mu t})^{j-n}.e^{-\mu n t} \tag{8}$$

Thus,

$$P^1(n) = p_n^j(\tau) = {}^jC_n(1 - e^{-\mu.\tau})^{j-n}e^{-j\mu\tau} \tag{9}$$

Rest of the values of $P^2(n)$, $P^3(n)$ etc. are calculated (as needed) by the use of theorem 3, described in the following part. Now before we prove two main theorems, we prove two simple lemmas as follows:

**Lemma 1**: For integer *j*,

$$\sum_n n^j.C_n.x^n = j.x(1+x)^{j-1} \tag{10}$$

**Proof** : We have from Binomial theorem that

$$(1+x)^j = \sum_n {}^jC_n . x^n \tag{11}$$

Since, the series on the right-hand side of (11) converges for all finite values of '$x$', differentiating both sides with respect to '$x$' and then multiplying '$x$' on both sides, we get the required result.

**Lemma 2**: For integer $j$,

$$\sum_n n^2 . {}^jC_n x^n = j.x(1+x)^{j-1} + j.(j-1).x^2.(1+x)^{j-2} \tag{12}$$

**Proof**: The result immediately follows by differentiating both sides of (10) with respect to '$x$', and then multiplying '$x$' on both sides.

Now we develop a recursive relation for $P^{(k)}(n)$ as follows:

**Theorem 3**: The probability distributions $P^{(k)}(n)$ s are recursively given as:

$$P^{(k+1)}(n) = \sum_i P^{(k)}(i) {}^{q+i}C_n (1-e^{-\mu.\tau})^{q+i-n} e^{-n.\mu.\tau} \quad \text{for } k = 0,1,2,\ldots \tag{13}$$

**Proof**: As per our notation, $P^{(k+1)}(n)$ denotes the probability that at the instant $(k+1).\tau - 0$ (just before the next node deployment), there are '$n$' number of existing nodes. Probability of this event is the aggregate of the following mutually exclusive and exhaustive events: $E^{(k)}(i) \cap F^{(q+i)}(n)$, where $E^{(k)}(i)$ denotes the event that, at the instant $k.\tau.-0$, there were '$i$' number of existing nodes; so $P(E^{(k)}(i)) = P^{(k)}(i)$, and $F^{(q+i)}(n)$ denotes the event that, following the Markovian death-dynamics in the interval ($k.\tau$, $(k+1)\tau$) and starting with $(q + i)$ number of nodes, there are '$n$' number of existing nodes at the instant $(k+1)\tau - 0$. It should be remembered that as the death-dynamics equation is time-homogeneous, there is no implicit dependence of $F^{(q+i)}(n)$ on $k.\tau$. This way, we obtain the following relation:

$$P^{(k+1)}(n) = P(E^{(k+1)}(n)) = \sum_i P(E^{(k)}(i) \cap F^{(q+i)}(n)) \tag{14}$$

But because the events $E^{(k)}(i)$ and $F^{(q+i)}(n)$ are independent, we obtain

$$P^{(k+1)}(n) = \sum_i P(E^{(k)}(i)) P(F^{(q+i)}(n)) \tag{15}$$

From (9),

$$P(F^{(j)}(n)) = p_n^j(\tau) = {}^jC_n . (1-e^{-\mu t})^{j-n} . e^{-\mu n t} \tag{16}$$

Thus from (15) and (16) we obtain,

$$P^{(k+1)}(n) = \sum_{i} P^{(k)}(i)\,{}^{q+i}C_n(1-e^{-\mu.\tau})^{q+i-n}e^{-n.\mu.\tau}\,,\text{ for } k=0,1,2,\ldots$$

(Proved)

Theorem (1) gives the important formula for $P^{(k)}(n)$ upon which, rest of the development depends. As a visual aid, we have calculated $P^{(7)}(n), P^{(8)}(n), P^{(9)}(n), P^{(10)}(n)$ using Matlab and plotted the respective distribution in Figure (1).

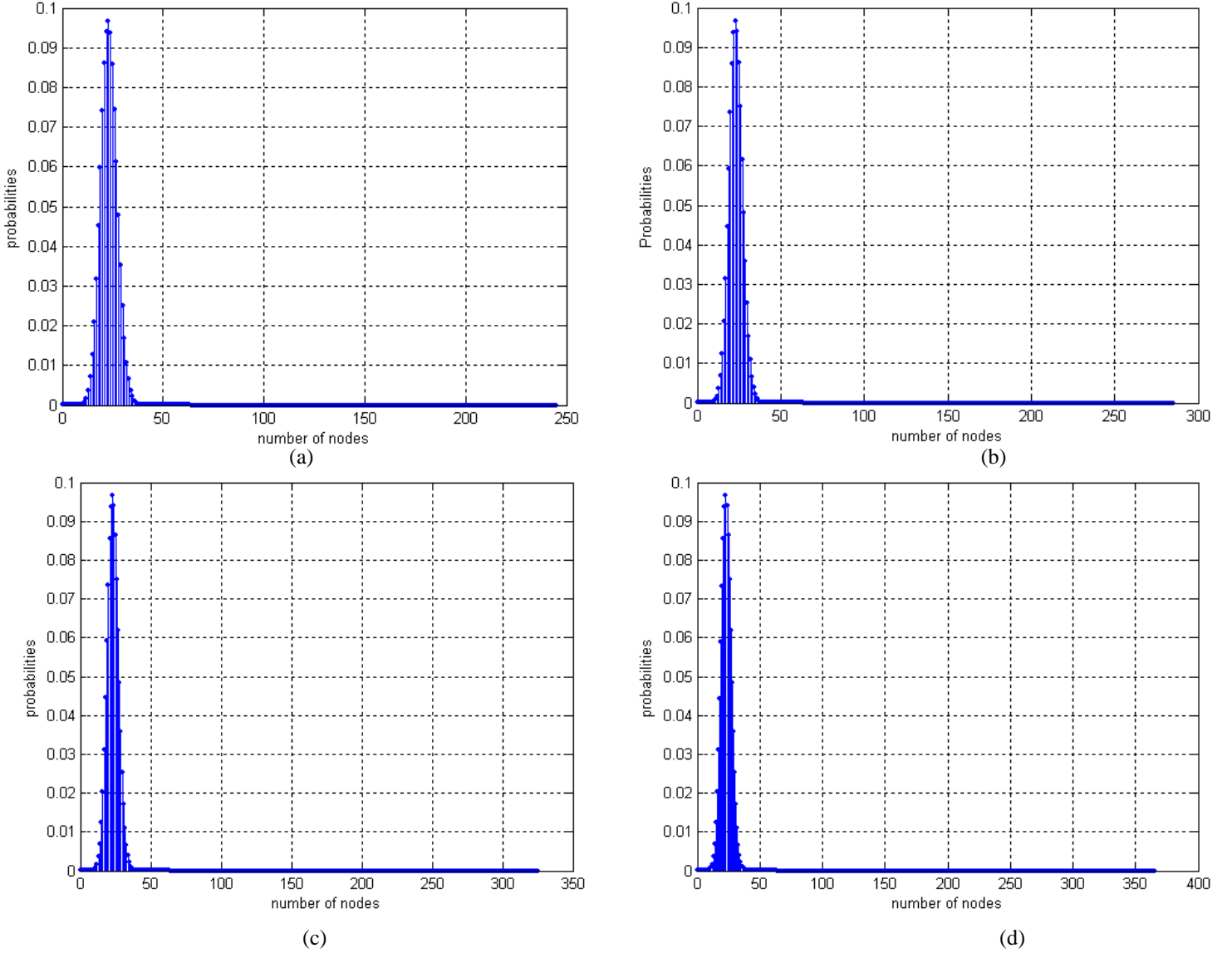

**Figure 1:** Probability distribution of $X^{(k)}$ for k=7 (a), 8 (b), 9 (c) and 10 (d)
(with $j=5, \mu\tau=1, r=23, q=40$);

Note the high-concentration of the distribution around the mean with increasing time. This property is exploited in section (4) in finding the optimal number of node re-deployment.

From the above plots, it appears intuitively clear that both the mean and variance of the distribution $P^{(k)}(n)$ tends to a steady-state value as $k \to \infty$. These two properties are proved rigorously in Theorems (4) and (5).

**Theorem 4:** The sequence of expectation of the stochastic process $X^{(k)}$ (denoted by $m^{(k)}$) converges to

$$\frac{q}{e^{\mu\tau}-1} \text{ as } k \to \infty$$

**Proof**: From Theorem 3, we have,

$$P^{(k+1)}(n) = \sum_i P^{(k)}(i)\, {}^{q+i}C_n (1-e^{-\mu\tau})^{q+i-n} e^{-n\mu\tau}, \text{ for } k = 0,1,2,\ldots$$

Now by definition,

$$m^{(k+1)} = \sum_n n.P^{(k+1)}(n)$$

$$= \sum_n n . \sum_i P^{(k)}(i)\, {}^{q+i}C_n (1-e^{-\mu\tau})^{q+i-n} e^{-n\mu\tau}$$

$$= \sum_i (1-e^{-\mu\tau})^{q+i} . P^{(k)}(i) \sum_n n \, {}^{q+i}C_n \frac{1}{(e^{\mu\tau}-1)^n} \quad \text{(Interchanging the order of summation)}$$

$$= \sum_i (1-e^{-\mu\tau})^{q+i} . P^{(k)}(i)(q+i) . \frac{1}{e^{\mu\tau}-1} . (1+\frac{e^{-\mu\tau}}{1-e^{-\mu\tau}})^{q+i-1} \quad \text{(Applying Lemma 1)}$$

$$= \frac{1-e^{-\mu\tau}}{e^{\mu\tau}-1} \sum_i (q+i).P^{(k)}(i)$$

$$= qe^{-\mu\tau} \sum_i P^{(k)}(i) + e^{-\mu\tau} \sum_i i.P^{(k)}(i) \qquad (17)$$

But $P^{(k)}(i)$ being a probability distribution, $\sum_i P^{(k)}(i) = 1$ and also, $\sum_i i.P^{(k)}(i) = m^{(k)}$ (by definition). So we obtain the difference equation for $m^{(k)}$ as follows:

$$m^{(k+1)} = qe^{-\mu\tau} + e^{-\mu\tau}.m^{(k)} \qquad (18)$$

Since there were initially ($k = 0$) $j$ nodes, we get $m^{(0)} = j$. So, taking Z-transform [38, 39] of both sides of (3), we obtain

$$z(M(z) - m(0)) = qe^{-\mu\tau} \frac{z}{z-1} + e^{-\mu\tau} M(z)$$

which gives, 

$$M(z) = \frac{j.z}{z-e^{-\mu\tau}} + \frac{qe^{-\mu\tau}.z}{(z-1)(z-e^{-\mu\tau})}$$

Now only one pole of $M(z)$ is on the unit circle ($|z|=1$) and rest of the poles are within the unit circle (since $e^{-\mu\tau} < 1$ for all positive $\mu$ and $\tau$), the sequence $m^{(k)}$ will converge to a non-zero steady-state value [34]. The steady-state value of $m^{(k)}$ as $k \to \infty$ is obtained using the final value theorem [33]:

$$\lim_{k \to \infty} m^{(k)} = \lim_{z \to 1}(z-1).M(z)$$

$$= \frac{q}{e^{\mu\tau} - 1} \quad \text{(Proved)}$$

**Theorem 5**: The sequence of variance of the stochastic process $X^{(k)}$ (denoted by $Var^{(k)}$) converges to the steady state value $\frac{q}{2}\cos ech(\mu\tau)$ as $k \to \infty$

**Proof:** From the definition of variance, it follows that:

$$Var^{(k+1)} = E((X^{(k+1)})^2) - (m^{(k+1)})^2, \quad \text{(Here E denotes Expectation)}$$

$X^{(k)}$ denotes the stochastic process under consideration as defined previously.
Now,

$$E((X^{(k+1)})^2) = \sum_n n^2.P^{(k+1)}(n)$$

$$= \sum_n n^2. \sum_i P^{(k)}(i).^{q+i}C_n(1-e^{-\mu\tau})^{q+i-n} e^{-n\mu\tau}$$

$$= \sum_i P^{(k)}(i).(1-e^{-\mu\tau})^{q+i}. \sum_n n^2.^{q+i}C_n. \frac{1}{(e^{\mu\tau}-1)^n} \quad \text{(Interchanging the order of summation)}$$

$$= \sum_i P^{(k)}(i).[e^{-\mu\tau}(q+i) + e^{-2\mu\tau}.(q+i).(q+i-1)] \quad \text{(Applying Lemma 2)} \quad (19)$$

Using (18) we get,

$$Var^{(k+1)} = \sum_i P^{(k)}(i)[e^{-\mu\tau}(q+i) + e^{-2\mu\tau}(q+i)(q+i-1)] - (qe^{-\mu\tau} + e^{-\mu\tau}m^{(k)})^2$$

After some algebraic manipulations the difference equation for $Var^{(k+1)}$ becomes

$$Var^{(k+1)} = e^{-2\mu\tau}Var^{(k)} + (e^{-\mu\tau} - e^{-2\mu\tau})q + (e^{-\mu\tau} - e^{-2\mu\tau})m^{(k)} \quad (20)$$

since initially, there were $j$ number of nodes with probability 1, we have

$$Var^{(0)} = 0 \quad (21)$$

Now taking Z-transform of (20) and taking (21) into account, we get

$$zVar(z) = e^{-2\mu\tau}Var(z) + (e^{-\mu\tau} - e^{-2\mu\tau})q\frac{z}{z-1} + (e^{-\mu\tau} - e^{-2\mu\tau})M(z)$$

or, $$Var(z) = q(e^{-\mu\tau} - e^{-2\mu\tau}) \cdot \frac{z}{(z-1)(z-e^{-2\mu\tau})} + \frac{(e^{-\mu\tau} - e^{-2\mu\tau})}{(z-e^{-2\mu\tau})} \cdot M(z)$$

Since only one of the pole of $Var(z)$ is on the unit circle and the rest of the poles are within the unit circle ($e^{-\mu\tau} < 1$ and $e^{-2\mu\tau} < 1$ for all positive $\mu$ and $\tau$) and only one pole is on the unit circle, the sequence $Var^{(k)}$ converges to a non-zero steady state value. The steady-state value of $Var^{(k)}$ is obtained by the Final Value Theorem as follows:

$$\lim_{k \to \infty} Var^{(k)} = \lim_{z \to 1}(z-1).Var(z)$$

$$= \frac{q(e^{-\mu\tau} - e^{-2\mu\tau})}{(1 - e^{-2\mu\tau})} + \frac{(e^{-\mu\tau} - e^{-2\mu\tau})}{(1 - e^{-2\mu\tau})} \cdot \lim_{z \to 1}(z-1).M(z)$$

$$= \frac{qe^{-\mu\tau}}{1 - e^{-2\mu\tau}} \qquad \text{[Using Theorem 4]}$$

A simple manipulation shows that $\lim_{k \to \infty} Var^{(k)} = \frac{q}{2} \cos ech(\mu.\tau) = \text{Constant}$. (Proved)

**Remark 1:** It appears that the steady-state values of the mean and variance of the existing number of nodes are independent of the initial number of nodes present in the network.

**Remark 2:** Since we are adding a finite ($q$) number of nodes at a periodic time-interval of $\tau$, intuitively there may exist every possibility of overflowing the sensor-network operational area with infinitely large number of nodes and thus rendering the system to be very inefficient and even non-operative(because of strong electromagnetic interference between two neighboring nodes due to their proximity). Theorem 4 ensures that on the average this situation does not happen. On the other hand Theorem 5 ensures that variability of the number of node is also bounded, thus ensuring the reliability of the system (since it meets the minimum number of nodes constraint, see section 4.). Thus these two theorems taken together provide the theoretical background for smooth functioning of the sensor-network system. The convergence characteristics of the mean and variance of existing number of nodes have been shown in Figure 2 for two different initial conditions through a MATLAB simulation.

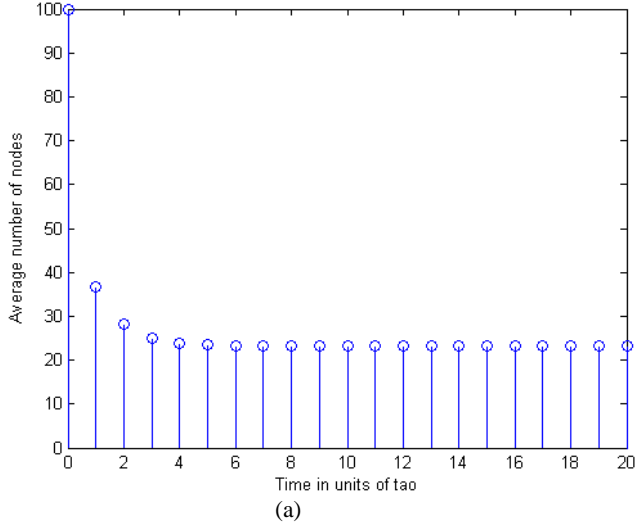
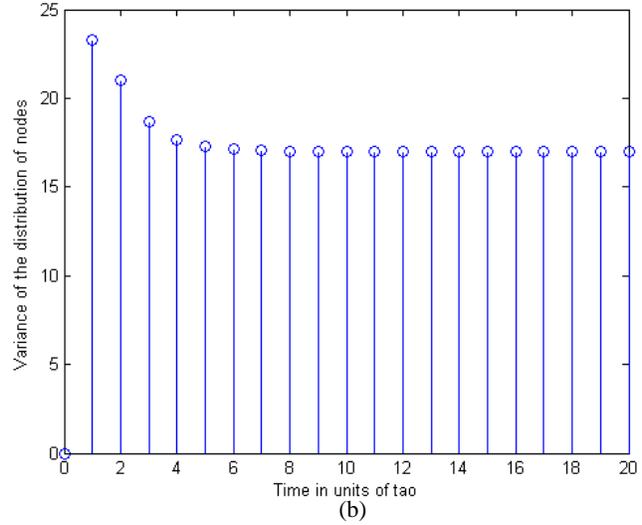
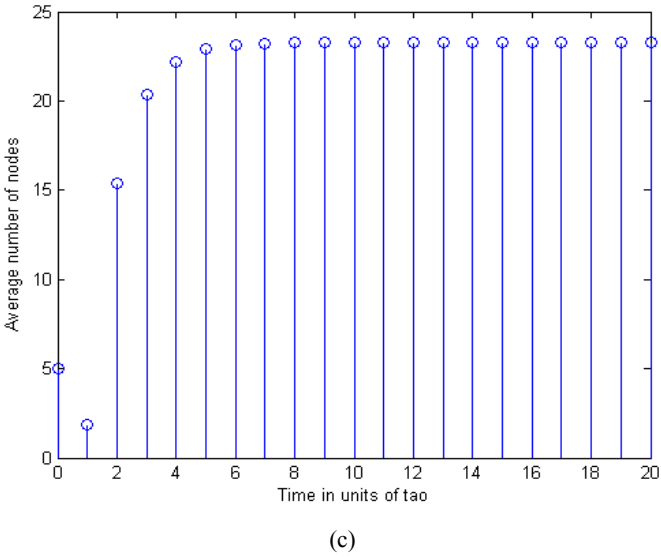
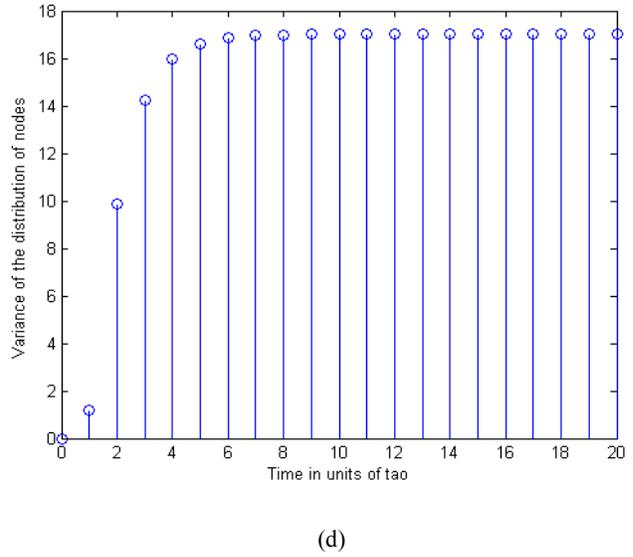

**Figure 2**: The convergence characteristics of the mean and variance of existing number of nodes with respect to time in units of $\tau$. For (a) and (b), initial number of nodes =100, for (c) and (d) initial number of nodes =5. Still both of them converge to the same steady-state values.

## 4. Node Deployment Schedule for Maintaining Optimal Connectivity in WSN

It must be noted that the parameters $q$ and $\tau$ are still kept unspecified. Theorem (4) and (5) holds for any finite values of $q$ and $\tau$. In this section we will find the optimum values for both of them. Suppose the minimum number of nodes for maintaining the connectivity be $r$ and we deploy $q$ sensor nodes at time interval of $\tau$. Initially the total number of nodes was $j$ with $j > r$. We want to find out the optimum values of $\tau$ and $q$. First we calculate the time-averaged cost for keeping the WSN operational and subsequently minimize it to find the optimized values. The cost comes from two factors

1. A fixed cost for node-deployment each time (network initialization cost, transport cost for re-deployment and subsequent overhead etc.)
2. Cost for deploying $q$ number of nodes after each $\tau$ time interval.

Let us consider a large time interval $T = k.\tau$ where $k$ is a positive integer and $k >> 1$.

So the total cost in time interval $T$ (assuming a linear cost function), is given by

$$C = \alpha.q.k + \beta.k \qquad (22)$$

Where $\alpha, \beta$ (>0) denotes the relative weights of costs due to factors (2) and (1) (mentioned above) respectively. Before we derive the expression for $\frac{C}{T}$, it is necessary to find an expression for $q$ which will ensure connectivity for all time.

**Expression of $q$ for maintaining minimum connectivity:**

From Chebyshev's inequality, we know that for any random variable X, and for $t > 0$

$$\Pr[|X - E(X)| \geq t] \leq \frac{Var(X)}{t^2}, \qquad (23)$$

with the equality sign holding only in a uniform distribution.

Again we have that,

$$\Pr[|X - E(X)| \geq t] = \Pr[X - E(X) \geq t] + \Pr[E(X) - X \geq t] > \Pr[E(X) - X \geq t] \qquad (24)$$

Thus from equation (23), we obtain

$$\Pr[E(X) - X \geq t] < \frac{Var(X)}{t^2} \qquad (25)$$

From the central tendency of $X^{(k)}(n)$ as depicted in Figure (1), it is clear that the result in (25) gives a more than necessary conservative estimate of lower tail-probabilities of $X^{(k)}(n)$ as $k \to \infty$. So we take a moderate value of the upper-bound probability, say 0.1. Then we have $t \approx 3\sigma$. This implies that,

$$\Pr[X^{k)} < E(X^{(k)}) - 3\sigma] < 0.1 \text{ as } k \to \infty$$

So we may assume the lowest "possible" value of $q$ (with a significant probability) to be equal to $\mu - 3\sigma$.
To maintain node-connectivity, if the minimum number of nodes required $= r$, then we need

$$\frac{q}{e^{\mu\tau} - 1} - 3\sqrt{\frac{qe^{-\mu\tau}}{1 - e^{-2\mu\tau}}} \geq r \qquad (26)$$

(We put the values of $\mu$ and $\sigma$ from Theorem (4) and (5) respectively).

This condition ensures that the WSN remains operational for the entire future time with a sufficiently high probability. Changing the inequality to equality for the optimal case and solving the resulting quadratic equation for $q$, we get $q$ to be equal to:

$$q = \frac{1}{2}(e^{\mu\tau}-1)(2r+\frac{9e^{\mu\tau}}{e^{\mu\tau}+1}) + \frac{1}{2}\sqrt{(e^{\mu\tau}-1)^2(2r+\frac{9e^{\mu\tau}}{e^{\mu\tau}+1})^2 - 4r^2(e^{\mu\tau}-1)^2} \quad (27)$$

(The other solution of the quadratic equation does not satisfy equation (26)).

Equation (27) gives the required value of $q$, maintaining connectivity with a high probability.

Now we turn our attention to the calculation of the time-averaged cost function. Putting the value of $q$ from (27) and $k = \frac{T}{\tau}$ in (22) we obtain,

$$\frac{C}{T} = \xi = \frac{\alpha}{2\tau}(e^{\mu\tau}-1)(2r+\frac{9e^{\mu\tau}}{e^{\mu\tau}+1}) + \frac{1}{2\tau}\sqrt{(e^{\mu\tau}-1)^2(2r+\frac{9e^{\mu\tau}}{e^{\mu\tau}+1})^2 - 4r^2(e^{\mu\tau}-1)^2} + \frac{\beta}{\tau}$$

Here $\frac{C}{T}$ denotes the time-average maintenance cost for the network for time $T$. If we take $T \to \infty$ then we get the large-scale time average maintenance cost of the network

$$\frac{C}{T} = \xi = \frac{\alpha}{2\tau}(e^{\mu\tau}-1)(2r+\frac{9e^{\mu\tau}}{e^{\mu\tau}+1}) + \frac{1}{2\tau}\sqrt{(e^{\mu\tau}-1)^2(2r+\frac{9e^{\mu\tau}}{e^{\mu\tau}+1})^2 - 4r^2(e^{\mu\tau}-1)^2} + \frac{\beta}{\tau} \quad (28)$$

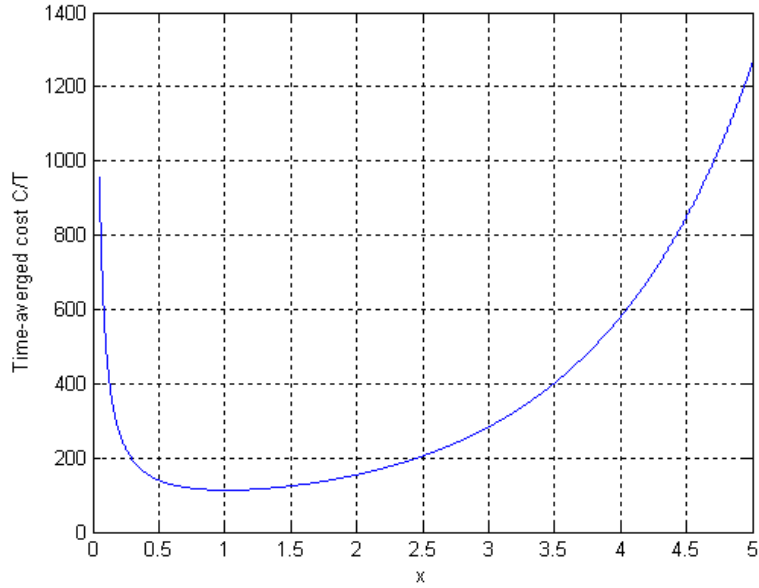

**Figure 3:** A plot of the Cost function (with $\beta = 2.\alpha.r$ and $r = 23$)

To minimize ξ, we set $\dfrac{d\xi}{d\tau} = 0$ and find the optimum $\tau$.

$\dfrac{d\xi}{d\tau} = 0$ Implies that

$$\alpha.x.\{2re^x + 9\dfrac{e^x(e^{2x}+2e^x-1)}{(e^x+1)^2}\} - \alpha\{2r(e^x-1)+9e^x\dfrac{(e^x-1)}{(e^x+1)}\} +$$

$$\alpha.x.\dfrac{(e^x-1)\{(2r+\dfrac{9e^x}{e^x+1})(2re^x+9\dfrac{e^x(e^{2x}+2e^x-1)}{(e^x+1)^2})-4r^2e^x\}}{\sqrt{\{2r(e^x-1)+9\dfrac{e^x(e^x-1)}{(e^x+1)}\}^2 - 4r^2(e^x-1)^2}} - \alpha.\sqrt{\{2r(e^x-1)+9\dfrac{e^x(e^x-1)}{e^x+1}\}^2 - 4r^2(e^x-1)^2}$$

$$-2.\beta = 0,$$

where $x = \mu\tau$.

Since the values of α, β, and r are known, solution of this complicated transcendental equation may be readily obtained graphically. Thus x, hence the optimized time interval τ is known and we get the optimized values of $q$ (number of nodes re-deployed periodically at time interval τ) from equation (27). From the graph in Figure 3 it follows for $\mu\tau = 1, r = 23, \beta = 2.\alpha.r$ that the cost is minimized when $x \equiv \mu\tau = 1.0197$. In this case $q$ is calculated from equation (27) to be $\approx 70$.

## 5. Conclusions and Future Research

In this paper we have derived a novel probabilistic model for the periodic node deployment in a WSN. Through some rational assumptions, this article puts forward a mathematical model for maintaining QoS in WSN, namely connectivity and coverage. On the theoretical part, we have proved that the process of adding deterministic amount of nodes at a periodic time interval always results in convergent mean and variance of number of existing nodes. These theoretical results have been utilized to find an optimum value of the time-interval and amount of nodes re-deployed at that interval by using a linear cost function. Some properties arising out of the discrete-birth and continuous-death of the sensor-nodes have also been established.

The future direction of research would be to include the non-uniform spatial profile of the nodes. Since the central nodes and nodes near to the base stations carry more traffic (due to forwarding of other node's packets) than peripheral nodes, a more appropriate model for the nodes will involve death rates ($\mu_n's$) as a function of radial distance(*r*). Then the deployment of nodes will also be a function of *r*.

## References

1. E. H. Callaway, Jr., *Wireless Sensor Networks: Architectures and Protocols*, CRC Press, August 2003.
2. F. Zhao and L. Guibas, *Wireless Sensor Networks: An Information Processing Approach*, Morgan Kaufmann, 2004.
3. N. Bulusu and S. Jha, *Wireless Sensor Network: A Systems Perspective*, Artech House, July 2005.
4. C. Chong and S. Kumar, "Sensor Networks: Evolution, Opportunities, and Challenges," *Proc. IEEE*, vol. 91, no. 8, pp. 1247-1256, Aug. 2003.
5. 10 emerging technologies that will change your world, *IEEE Engineering Management Review*, 2004, 32(2): 20-30.
6. E. Shih, S.-H. Cho, N. Ickes, R. Min, A. Sinha, A. Wang, and A. Chandrakasan, "Physical layer driven protocol and algorithm design for energy-efficient wireless sensor networks", In *ACM Int'l Conf. on Mobile Computing and Networking (MobiCom)*, pages 272–287, 2001.
7. A. Woo and D. E. Culler, "A transmission control scheme for media access in sensor networks", In *ACM Int'l Conf. on Mobile Computing and Networking (MobiCom)*, pages 221–235, 2001.
8. W. Ye, J. Heidemann, and D. Estrin, "An energy-efficient MAC protocol for wireless sensor networks", In *IEEE INFOCOM*, pages 1567–1576, 2002.
9. D. Braginsky and D. Estrin, "Rumor routing algorithm for sensor networks", In *ACM Int'l Workshop on Wireless Sensor Networks and Applications (WSNA)*, 2002.
10. C. Intanagonwiwat, R. Govindan, D. Estrin, J. Heidemann, and F. Silva, "Directed Diffusion for Wireless Sensor Networking," *IEEE/ACM Trans. Networking*, vol. 11, no. 1, pp. 2-16, Feb. 2003.
11. W. R. Heinzelman, A. Chandrakasan, and H. Balakrishnan, "Energy-efficient communication protocols for wireless microsensor networks", In *Hawaii Int'l Conf. on Systems Science (HICSS)*, 2000.
12. H. Luo, Y. Liu, and S. K. Das, "Routing Correlated Data with Fusion Cost in Wireless Sensor Networks," *IEEE Trans. Mobile Computing*, vol. 5, no. 11, Nov. 2006.

13. D. Ganesan, R. Govindan, S. Shenker, and D. Estrin, "Highly resilient, energy efficient multipath routing in wireless sensor networks", *ACM Mobile Comput. and Commun. Review*, 5(4):11–25, Oct. 2001.
14. P. Bahl and V. N. Padmanabhan, "RADAR: An in-building RF-based user location and tracking system", In *IEEE INFOCOM*, pages 775–784, 2000.
15. N. Bulusu, J. Heidemann, and D. Estrin, "GPS-less low cost outdoor localization for very small devices", *IEEE Personal Commun.*, 7(5):28–34, Oct. 2000.
16. D. Nicules and B. Nath, "Ad-hoc positioning system (APS) using AoA", In *IEEE INFOCOM*, 2003.
17. A. Savvides, C.-C. Han, and M. B. Strivastava, "Dynamic fine-grained localization in ad-hoc networks of sensors", In *ACM Int'l Conf. on Mobile Computing and Networking (MobiCom)*, pages 166–179, 2001.
18. Y.-C. Tseng, S.-P. Kuo, H.-W. Lee, and C.-F. Huang, "Location tracking in a wireless sensor network by mobile agents and its data fusion strategies", In *Int'l Workshop on Information Processing in Sensor Networks (IPSN)*, 2003.
19. J. N. Al-Karaki and A. E. Kamal, "Routing Techniques in Wireless Sensor Networks: A Survey", *IEEE Wireless Communications*, pp. 6-28, December 2004.
20. I. F. Akyildiz, W. Su, Y. Sankarasubramaniam, and E Cayirci, "Wireless sensor networks: a survey", *Computer Networks*, Vol. 38, No. 4. (15 March 2002), pp. 393-422.
21. G. Pottie and W. Kaiser, "Wireless sensor networks", *Communications of the ACM*, vol. 43, no. 5, pp. 51–58, May 2000.
22. Z. Bojkovic and B. Bakmaz, "A survey on wireless sensor networks deployment", WSEAS Trans. On Communications, Vol. 7, No. 12, Dec. 2008, pp. 1172-1181.
23. J. Yick, B. Mukherjee, and D. Ghosal, "Wireless sensor network survey", *Computer Networks*, Vol. 52, No. 12. pp. 2292-2330, Aug. 2008.
24. D. P. Mehta, M. A. Lopez, L. Lin, "Optimal coverage paths in ad-hoc sensor networks", Proceedings of International Conference on Communications (ICC'2003): Vol 1, Anchorage, AK, USA. Piscataway, NJ, USA: IEEE, 2003, 507-511.
25. C-F Huang, and Y-C Tseng, "The Coverage Problem in a Wireless Sensor Network", *WSNA'03,* San Diego, California, USA, September 19, 2003.
26. A. Sekhar, B. S. Manoj, C. Siva Ram Murthy, "Dynamic coverage maintenance algorithms for sensor networks with limited mobility", Proc. of 3rd *International Conference on Pervasive Computing and Communications*, Mar 8-12, 2005, Piscataway, NJ, USA: IEEE Computer Society, 2005, 51-60.
27. S. Meguerdichian, F. Koushanfar, M. Potkonjak, and M. Srivastava, "Coverage Problems in Wireless Ad-Hoc Sensor Networks," IEEE Infocom 2001, Vol 3, pp 1380-1387, April 2001.
28. M. T. Thai, F. Wang, D. H. Du, and X. Jia, "Coverage problems in wireless sensor networks: designs and analysis", International *Journal of Sensor Networks,* 3( 3) , pp.191-200, Inderscience Publishers, 2008.
29. Y. Liu, H. Ngan, and L.M. Ni, "Power-Aware Node Deployment in Wireless Sensor Networks," Int'l *J. Distributed Sensor Networks*, vol. 3, pp. 225-241, Apr. 2007.
30. S. Megerian, F. Koushanfar, M. Potkonjak, and M. B. Srivastava, "Worst and best-case coverage in sensor networks", *IEEE Transactions on Mobile Computing,* .4(1): 84-92, 2005.
31. Z. Yan and B. Zheng, "A novel mathematical model for coverage in wireless sensor network", *Journal of China Universities of Posts and Telecommunications*, Volume 13, Issue 4, December 2006.
32. D. Wang, B. Xie, and D. P. Agrawal, Coverage and Lifetime Optimization of Wireless Sensor Networks with Gaussian Distribution, *IEEE Trans. on Mobile Comp.*, Vol. 7, No. 12, Dec. 2008.
33. G. Latouche and V. Ramaswami, *Introduction to Matrix Analytic Methods in Stochastic Modeling*, 1st edition, Chapter 1: *Quasi-Birth-and-Death Processes*; ASA SIAM, 1999.
34. S. N. Ethier and T. G. Kurtz, *Markov Processes: Characterization and Convergence*, John Wiley & Sons, 1986.

35. W. J. Stewart, *Introduction to the Numerical Solution of Markov Chains*, Princeton University Press, pp. 17–23, 1994.
36. P. Olofsson, *Probability, Statistics, and Stochastic Processes*, Wiley-Interscience, 2005.
37. A. Gut, *Probability: A Graduate Course*, Springer (Texts in Statistics), 2005.
38. R. E. Attar, *Lecture notes on Z-Transform*, Lulu Press, Morrisville NC, 2005.
39. K. Ogata, Discrete Time Control Systems 2nd Ed, Prentice-Hall Inc, 1995, 1987.